\documentclass[11pt,print,amsmath,amssymb,prd]{revtex4-2}
\usepackage{CJK}

\usepackage{graphicx}
\usepackage{dcolumn}
\usepackage{bm}

\usepackage[colorlinks, linkcolor=blue]{hyperref}
\usepackage{subcaption}
\usepackage{amsmath}
\usepackage{amsthm,amssymb,bbold}
\usepackage{array} 
\usepackage[dvipsnames]{xcolor}

\begin{document}
\begin{CJK*}{GB}{} 

\title{Supersolidity-related
phenomena in holographic superfluid: cnoidal wave}
\author{Peng Yang$^1$}\email{yangpeng18@mails.ucas.edu.cn}
\author{Yu Tian$^{1,2}$}\email{ytian@ucas.ac.cn}

\affiliation{$^1$School of Physical Sciences, University of Chinese Academy of Sciences, Beijing 100049, China}
\affiliation{$^2$Institute of Theoretical Physics, Chinese Academy of Sciences, Beijing 100190, China}

\begin{abstract}
The candidate of supersolid-like state, cnoidal wave, is investigated in finite temperature holographic superfluid model for the first time. We find by giving different source $a_x$ for the particle current density $\langle j^{x}\rangle$, different kinds of superflow states can be obtained numerically. And based on $a_x$, three limiting cases of cnoidal waves can be found easily as well as the general cnoidal wave solution. The chemical potential $\mu$ and particle current density $\langle j^x \rangle$, as functions of source $a_x$, are calculated for these superflow states and are compared with homogeneous superflow. We find as we increasing $a_x$, cnoidal waves will enter into the uniform limit, which reflects in the confluence of the behavior of $\mu(a_x)$ and $\langle j^x(a_x) \rangle$ between cnoidal wave and homogeneous superflow.  Based on the quasi-normal modes, we show the cnodial waves are always dynamically unstable due to the finite temperature, while the energetic stability is influenced by the particle current density. All these evidences manifest that the cnoidal wave can be seen as an \textcolor{black}{unstable excited states} with supersolidity in finite temperature superfluid system.

\end{abstract}

\maketitle
\tableofcontents
\end{CJK*}
\section{Introduction}

Symmetry occupies an indispensable and fundamental role in the language of physics and especially important in condensed matter physics, particle physics and so on. Usually, a system transforms from one phase into another phase, there must be accompanied with the spontaneous symmetry broken (SSB). For instance, the formation of solid is related with the spontaneously broken of continuous space translation symmetry, the  appearance of superfluid phase in quantum system is caused by the spontaneously broken of gauge $U(1)$ symmetry and the time translation symmetry spontaneously broken will produce the so-called time crystal. 

Supersolid, a matter of state that simultaneously possessing the superfluidiy and the long-range spatial crystalline order, has received continuous controversy due to the seeming contradiction between mobility of the superfluid and  the rigidity of the solid. However, more and more experiment researches \cite{2017Natur.543...91L,2017Natur.543...87L,2019PhRvL.122m0405T,2019Natur.574..386G,2019Natur.574..382T,2021Natur.596..357N} have confirmed the existence of supersolid or discovered the supersolidity in BEC system since \cite{2017Natur.543...91L}. Using Bragg reflection, \cite{2017Natur.543...91L} has detected a stripe phase with supersolid properties in a BEC system with the spin-orbit coupling. \cite{2017Natur.543...87L} realizes the formation of supersolid by introducing the cavity-mediated long-range interactions into BEC system. In dipolar quantum gas systems, the competition between dipolar-dipolar interaction and constant interaction will also contribute to the formation of supersolid  \cite{2019PhRvL.122m0405T,2019Natur.574..386G,2019Natur.574..382T,2021Natur.596..357N}. The rigidity of supersolid in dipolar BEC experiment is able to provide opportunity to observe the low-energy Goldstone mode \cite{2019Natur.574..386G,2019Natur.574..382T}, which will also be an strong evidence for the detected supersolid. The realization of two dimensional supersolid \cite{2021Natur.596..357N} make much more potential researches being possible, such as the excitation properties.

Theoretically, supersolids are predicted in varied zero-temperature BEC systems \cite{2002PhRvL..88m0402G,2005PhRvL..95l7205W,2010PhRvL.104s5302H,2010PhRvL.105p0403W,2016PhRvX...6b1026O} by adding the degree of freedom of interaction that the BEC suffered. For instance, the dipolar-dipolar interaction \cite{2002PhRvL..88m0402G}, the nearest-neighbor repulsion of the hard-core bosons on a lattice \cite{2005PhRvL..95l7205W}, the Rydberg interactions with Rydberg state \cite{2010PhRvL.104s5302H}, the spin-orbit interaction in spin-$1/2$ and spin-1 BEC \cite{2010PhRvL.105p0403W}, the nonlinear atom-light interactions (superradiant Rayleigh scattering) \cite{2016PhRvX...6b1026O} and so on. Besides, in finite temperature systems, supersolids is also investigated in holographic model \cite{2015JHEP...09..168K,2022JHEP...06..152B}. Generally speaking, to form the favored supersolid state, one indispensable role is the roton \cite{2003PhRvL..90y0403S}. In \cite{2003PhRvL..90y0403S}, roton is one of the excitation of dipolar BEC, whose spectrum in energy-momentum space will have a 'roton minimum' and there has a 'roton gap' to stimulate the roton. Changing the form of interaction of BEC system may reduce the 'roton gap'. Until the 'roton gap' becomes zero, the system will experience roton instability and lead to condensate collapses \cite{2003PhRvL..90y0403S}, which gives the signal of appearance of supersolid. The collapsed condensate will eventually settle down to the supersolid state via some stabilization mechanism, such as three-body repulsion \cite{2015PhRvL.115g5303L} and quantum fluctuation \cite{2015PhRvL.115o5302P}.

It is worth noting that, in addition to excavating supersolids from ground state of systems, one can also investigate the supersolidity-related phenomena from excited unstable states or even non-equilibrium states. Just like one can construct gray-soliton in zero-temperature superfluid system \cite{2019JHEP...05..167G}, but one can not preserve the shape of moving 'soliton' in finite temperature. It is an open question to find supersolids-like state in finite temperature non-equilibrium systems, since it seems there always exists dissipation. In the present paper, we find the cnoidal wave \cite{2011PMag...91.1007.,1971JLTP....4..441T,2021PhRvR...3a3143M} is such kind of supersolid-like state in finite temperature superfluid system with current.

Cnoidal wave, whose investigation can be traced back to 1895 \cite{2011PMag...91.1007.}, also as a nonlinear stationary solution of Gross-Pitaevskii equation \cite{1971JLTP....4..441T}, is born with spatially periodic structure and hence getting more and more theoretical attention \cite{1974JAM...367,2011JPhA...44B5201B,2015JDE...258.3607G,2017AMR...431,2008PhRvL.100f0401K,2009PhRvA..79f3616K,2012PhRvA..86b3602B,2021PhRvR...3a3143M}. The stability analysis have shown that the cnoidal waves are dynamical stable at zero temperature \cite{1974JAM...367,2011JPhA...44B5201B,2015JDE...258.3607G,2017AMR...431}. While, the finite temperature situation hasn't been made certain up to now, which is the first motivation of present paper, i.e., we use finite temperature holographic superfluid model to investigate the stability of the cnoidal waves. Actually, there are three significant limiting cases for the cnoidal wave, the dark-soliton limit, linear-wave limit and the uniform limit. Among these three limiting cases, the dark-soliton limit is most common \cite{2008PhRvL.100f0401K,2009PhRvA..79f3616K,2012PhRvA..86b3602B,2021PhRvR...3a3143M}. In one dimensional BEC ring under rotation system \cite{2008PhRvL.100f0401K,2009PhRvA..79f3616K}, there is a transition from metastable uniform BEC to dark-solitons by changing the strength of interaction or the frequency of rotation. In one dimensional finite-range interaction BEC system \cite{2012PhRvA..86b3602B}, finite-range interaction leads to the construction of soliton train when the velocity of the system exceeds a critical value. This soliton train is also the dark-soliton limiting cnoidal wave. In holographic superfluid model, there doesn't have research about cnoidal wave, which motivate us to find it. And we find that the different limiting cases of cnoidal wave can be easily obtained by introducing particle current density source $a_x$. The corresponding properties of chemical potential and total current density are also investigated. Recently, the supersolidity of cnoidal wave is investigated in dark-soliton and linear-wave limiting cases at zero temperature, which points that the cnoidal wave has dynamical stability and Landau instability \cite{2021PhRvR...3a3143M}. While in the present paper, we will show that the non-zero temperature will influence the dynamical stability of cnoidal wave and the non-zero particle current density will influence the energetic stability.

This paper is organized as follows. In Sec.~\ref{Holographic Model} we introduce the holographic superfluid model that we use. We introduce some types (homogeneous or bloch) of superflow solution in Sec.~\ref{superflow state} and we analysis the relationship between these superflow states and the source $a_x$ in holographic model. The quasi-normal mode (QNM) analysis is given by Sec.~\ref{Quasi-normal mode}. The numerical calculation are given in Sec.~\ref{result}, from which in Sec.~\ref{CW} the homogeneous-type superflow and bloch-type cnoidal wave are solved and the chemical potential $\mu(a_x)$ as well as the particle current density $\langle j^x (a_x)\rangle$ are calculated as functions of source $a_x$. It is worth noting that, except the uniform limiting cnoidal wave, both of dark-soliton limiting and linear-wave limiting cnoidal waves have spatial crystalline order. The stability analysis of these superflow states in finite temperature are investigated in Sec.~\ref{qnm} and we can see that the spatial crystalline order will lead to the band structure of QNM in spectrum space. Finally, in Sec.~\ref{sum} we give a summary.

\section{Holographic superfluid model}\label{Holographic Model}

The simplest holographic superfluid model we use is proposed in \cite{2008PhRvL.101c1601H} and the action is
\begin{equation}
S(\Psi,A_{\mu})=\frac{1}{16 \pi G_{4}} \int d^{4} x \sqrt{-g}\left(R+\frac{6}{L^{2}}+\frac{1}{e^{2}} \mathcal{L}_{M}(\Psi,A_{\mu})\right),
\end{equation}
in which, a complex scalar field $\Psi$ is coupled to a $U(1)$ gauge field $A_{\mu}$ in the $(3+1)$D gravity with a cosmological constant related to the AdS radius as $\Lambda=-3/{L^2}$. The $G_{4}$ is the gravitational constant in four dimensional spacetime. The first two terms in the parenthesis are the gravitational part of the Lagrangian with the Ricci scalar $R$ and the AdS radius $L$, and the third consists of all matter fields given by
\begin{equation}
\mathcal{L}_{M}(\Psi,A_{\mu})=-\frac{1}{4} F_{\mu \nu}F^{\mu \nu}-\left|D_{\mu} \Psi\right|^{2}-m^{2}|\Psi|^{2}.
\end{equation}

The system we considered is the superfluid couples with a thermal bath with a constant temperature. So we take the probe limit \cite{2008PhRvL.101c1601H}, i.e., ${e}\to\infty$, to fix the background spacetime as the standard \textcolor{black}{Schwarzschild-AdS black brane} to provide a constant temperature with metric
\begin{equation}
ds^{2}=\frac{L^{2}}{z^{2}}\left(-f(z)dt^{2}+\frac{1}{f(z)}dz^{2}+dx^{2}+dy^{2}\right),\qquad f(z)=1-\frac{z^{3}}{z_{H}^{3}}.
\end{equation}
\textcolor{black}{Here $z_H$ is the radius of the black brane horizon.} The Hawking temperature of this black brane is $T=\frac{3}{4\pi z_{H}}$, which is also the temperature of the boundary system by the holographic dictionary. Furthermore, due to the existence of the bulk black hole the boundary field theory will intrinsically have dissipation, since there will be energy flow absorbed into the horizon \cite{2013Sci...341..368C,2019arXiv191201159T,2020JHEP...02..104L} \textcolor{black}{during dynamic processes}. For numerical simplicity, we set $L=1=z_{H}$.

The equations of motion for $A_{\mu}$ and $\Psi$ are
\begin{equation}\label{A_mu}
\frac{1}{\sqrt{-g}} \partial_{\mu}\left(\sqrt{-g} F^{\mu \nu}\right)=i\left(\Psi^{*} D^{\nu} \Psi-\Psi\left(D^{\nu} \Psi\right)^{*}\right).
\end{equation}
\begin{equation}\label{psi}
\frac{1}{\sqrt{-g}} D_{\mu}\left(\sqrt{-g} D^{\mu} \Psi\right)-m^{2} \Psi=0,
\end{equation}

A trivial solution with $\Psi=0$ that can be easily obtained from the above equations. As we increase the chemical potential $\mu$, there will be a critical chemical potential $\mu_{c}$ above which a solution with nonzero $\Psi$ appears, indicating the break of $U(1)$ symmetry and the formation of superfluid condensate. Hereafter, we will focus on the case with $\mu=4.5>\mu_{c}$, i.e., the superfluid phase of the boundary field theory.

For simplicity, we will consider the system is homogeneous in $y$ direction, so the static equations are
\begin{equation}\label{static Jx}
z^{2}\left(\partial_{z} f \partial_{z} A_{x}+f \partial_{z}^{2} A_{x}\right)=i\left(\Psi^{*} \partial_{x} \Psi-\Psi \partial_{x} \Psi^{*}-2 i \Psi A_{x} \Psi^{*}\right),
\end{equation}
\begin{equation}\label{static Jz}
-z^{2} \partial_{x} \partial_{z} A_{x}=i\left(\Psi^{*} \partial_{z} \Psi-\Psi\partial_{z} \Psi^{*}\right),
\end{equation}
\begin{equation}\label{static Jt}
z^{2}\left(f \partial_{z}^{2} A_{t}+\partial_{x}^{2} A_{t}\right)=2 A_{t} \Psi^{*} \Psi,
\end{equation}
\begin{equation}\label{static Psi}
\frac{z^{2}}{f} A_{t}^{2} \Psi+z^{4} \partial_{z}\left(z^{-2}f\right) \partial_{z} \Psi+z^{2} f \partial_{z}^{2} \Psi+z^{2}\left(\partial_{x}-i A_{x}\right)^{2} \Psi-m^{2} \Psi=0.
\end{equation}

From the asymptotic analysis of all these static fields we know
\begin{equation}
\Psi \sim z^{d-\Delta_{+}} \Psi_{+}+z^{d-\Delta_{-}} \Psi_{-}+\ldots,
\end{equation}
\begin{equation}\label{At}
A_{t} \sim \mu-z^{d-2} \rho+\ldots,
\end{equation}
\begin{equation}
A_{x} \sim a_{x}-z^{d-2} \langle j^{x}\rangle+\ldots,
\end{equation}
where $\Delta_{\pm}=\frac{d \pm \sqrt{d^{2}+4 m^{2}}}{2}$ with $d=3$ in our case. The holographic dictionary tells us that with one of $\Psi_{\pm}$ being the source the other will be the related response, $\mu$ is the chemical potential, $\rho$ the \textcolor{black}{(conserved)} particle number density and $a_{x}$ the source of the \textcolor{black}{particle current density} $\langle j^{x}\rangle$ in the boundary field theory.  It is convenient to choose $m^{2}=-2$ to make all calculations easier and we use standard quantization, in which case
\begin{equation}
	\Psi \sim z \Psi_{0}+z^{2} \langle \mathcal O \rangle+O\left(z^{3}\right)=:z \psi,
\end{equation}
with $\Psi_{0}$ the source and $\langle \mathcal O \rangle$ as the response. Here we define a new bulk field $\psi$.

\subsection{Superflow states with homogeneous-type and bloch-type}\label{superflow state}

\begin{figure}
\centering
\begin{subfigure}[b]{0.45\textwidth}
\includegraphics[width = \textwidth]{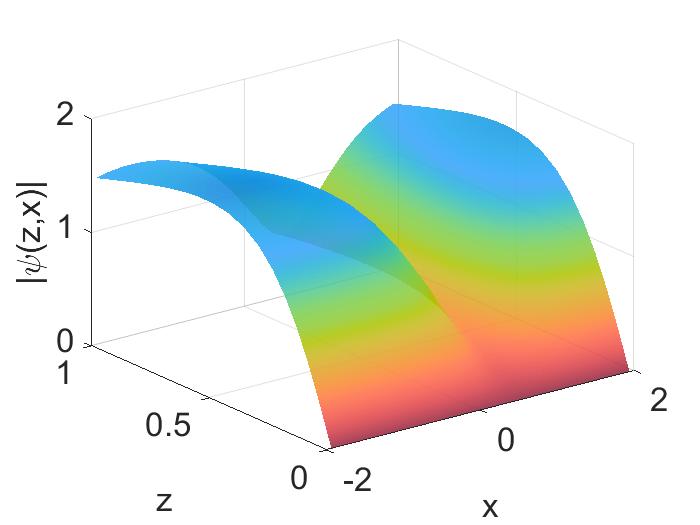}
\caption{}
\label{CW_bulk}
\end{subfigure}
\begin{subfigure}[b]{0.45\textwidth}
\includegraphics[width = \textwidth]{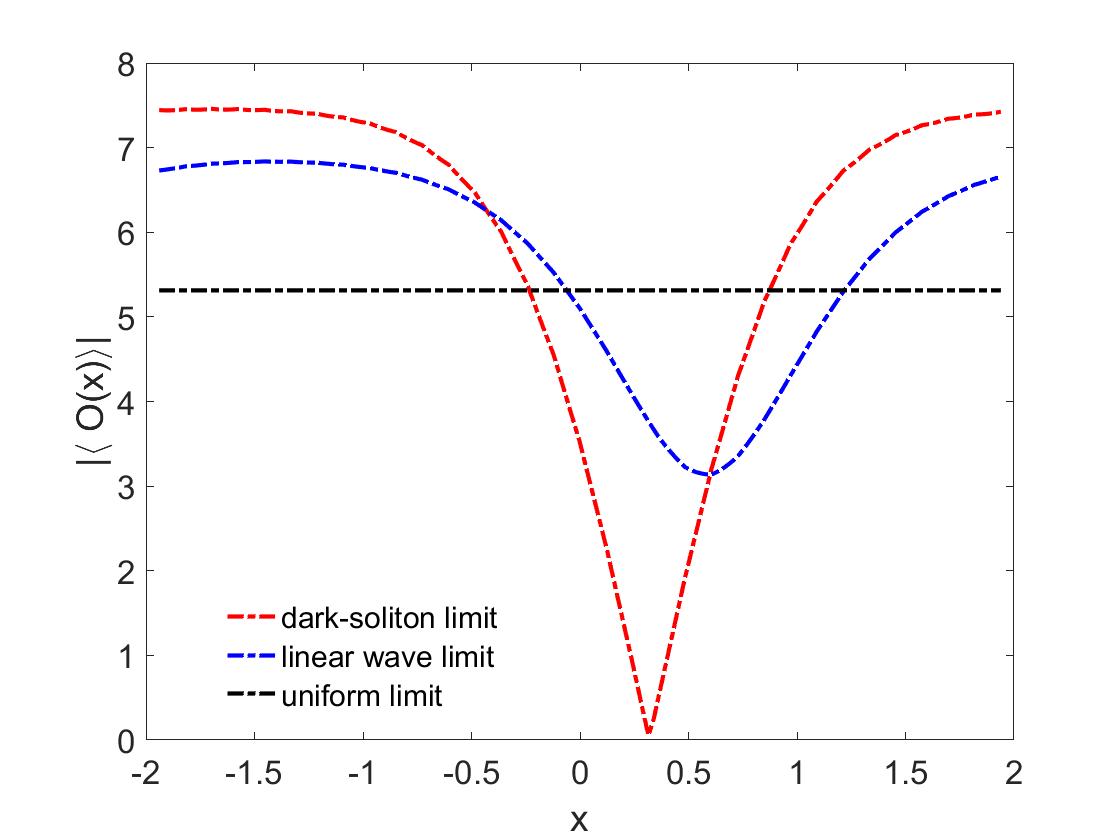}
\caption{}
\label{CW_bdry}
\end{subfigure}
\caption{Fig.\ref{CW_bulk} shows bulk field $|\psi(z,x)|$ of general form of cnoidal wave ($a_x=\frac{5\pi}{8}$). Fig.\ref{CW_bdry} shows the profile of $|\langle \mathcal O(x) \rangle|$ in dark-soliton limit (red line, $a_x=\frac{\pi}{4}$), linear-wave limit (blue line, $a_x=\frac{7 \pi}{8}$) and uniform limit (black line, $a_x={\pi}$), respectively.}
\label{states}
\end {figure}

Consider the system that without source $\Psi_{0}$ and $a_x$, $\langle \mathcal O \rangle$ is order parameter for boundary superfluid system. The homogeneous superflow states corresponding to a constant value of $|\langle \mathcal O \rangle|$ and a phase $\theta(x)=k x$. While if the system has spatial crystalline order, a set of bloch states $\langle \mathcal O(x) e^{i k x} \rangle$,  can describe the spatial crystalline ordered superflow states. In such case, $\langle \mathcal O(x) \rangle$ is periodic and usually is a complex field. So without loss of generality, we can separate it with real and imaginary parts and this is equal to set $\psi=(\psi_R+i \psi_I) e^{i k x}$. Substituting the above definition into the static equations of motion \eqref{static Jx}-\eqref{static Psi}, we get the following five differential equations:
\begin{equation}\label{Jx}
\partial_{z} f \partial_{z} A_{x}+f \partial_{z}^{2} A_{x}=-2\left(\left(\psi_{R} \partial_{x} \psi_{I}-\psi_{I} \partial_{x} \psi_{R}\right)+\left(k-A_{x}\right)\left(\psi_{R}^{2}+\psi_{I}^{2}\right)\right),
\end{equation}
\begin{equation}\label{Jz}
\partial_{x} \partial_{z} A_{x}=2\left(\psi_{R} \partial_{z} \psi_{I}-\psi_{I} \partial_{z} \psi_{R}\right),
\end{equation}
\begin{equation}\label{Jt}
f \partial_{z}^{2}A_{t}+\partial_{x}^{2}A_{t}=2 A_{t}\left(\psi_{R}^{2}+\psi_{I}^{2}\right),
\end{equation}
\begin{equation}\label{Psi a}
\left({f^{-1}}A_{t}^{2}-z-\left(A_{x}-k\right)^{2}\right) \psi_{R}+\partial_{z}\left(f \partial_{z} \psi_{R}\right)+\left(\partial_{x}^{2} \psi_{R}-2\left(k-A_{x}\right) \partial_{x} \psi_{I}+\psi_{I} \partial_{x} A_{x}\right)=0,
\end{equation}
\begin{equation}\label{Psi b}
\left({f^{-1}} A_{t}^{2}-z-\left(A_{x}-k\right)^{2}\right) \psi_{I}+\partial_{z}\left(f \partial_{z} \psi_{I}\right)+\left(\partial_{x}^{2} \psi_{I}+2\left(k-A_{x}\right) \partial_{x} \psi_{R}-\psi_{R} \partial_{x} A_{x}\right)=0.
\end{equation}

The $U(1)$ symmetry of the system allows us re-define the field $\tilde A_x=A_x-k$ (Hereafter, we will use $A_x$ as the new field $\tilde A_x.$), then all the superflow states mentioned above can be obtained by giving source $a_x=-k$. This is the spirit of the holographic dictionary, i.e., $a_x$ is the source of the particle current density $\langle j^{x}\rangle$. 


\subsection{Quasi-normal mode of superflow states}\label{Quasi-normal mode}
\begin{figure}
\centering
\begin{subfigure}[b]{0.45\textwidth}
\includegraphics[width = \textwidth]{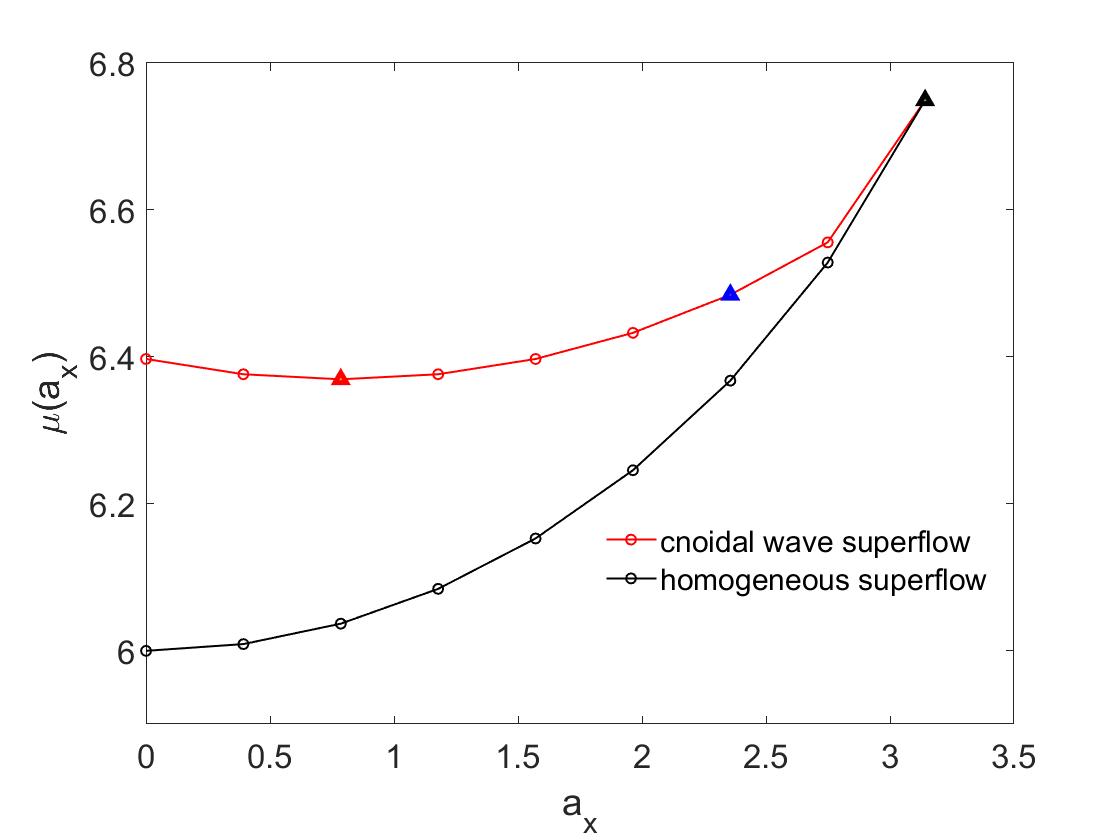}
\caption{}
\label{Band_CW}
\end{subfigure}
\begin{subfigure}[b]{0.45\textwidth}
\includegraphics[width = \textwidth]{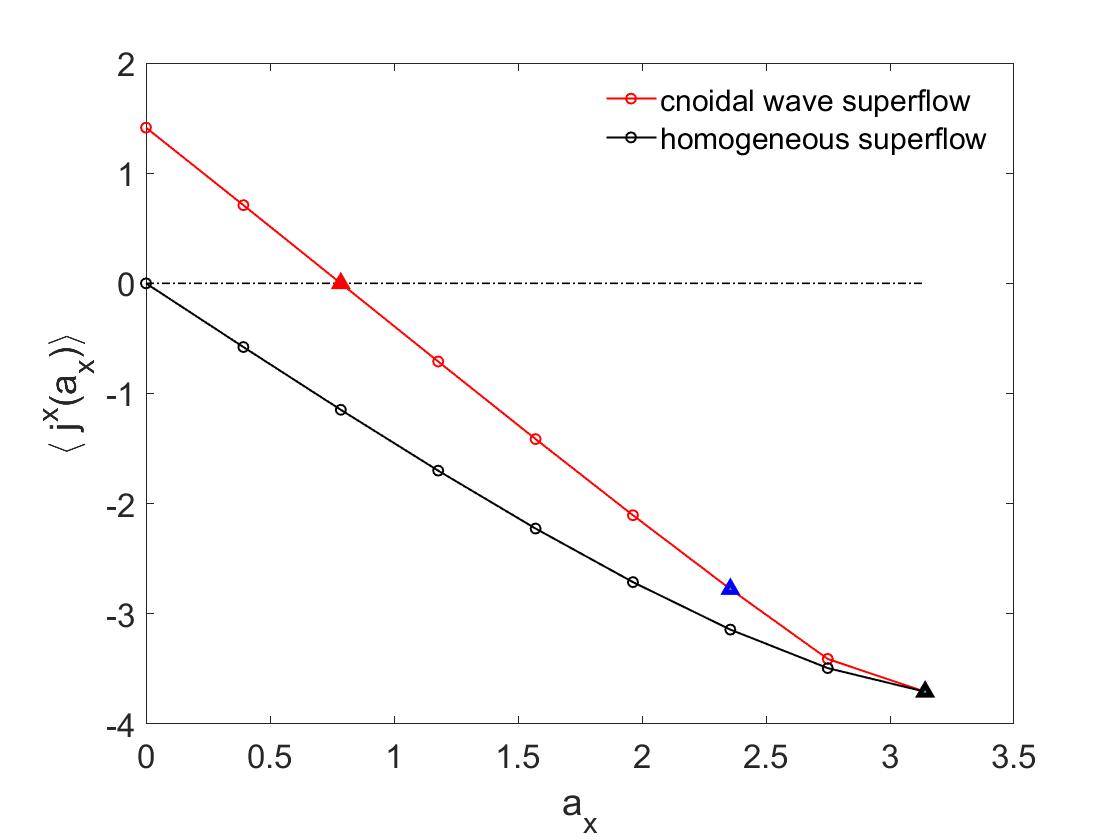}
\caption{}
\label{jx_CW}
\end{subfigure}
\caption{Fig.\ref{Band_CW}, Fig.\ref{jx_CW} show the chemical potential $\mu$ and total current density $\langle j^x \rangle$ as a function of source $a_x$, respectively. The red line stands for the cnoidal wave superflow and the black line stands for the homogeneous superflow. The filled triangle with red (dark-soliton limit), blue (linear-wave limit) and black (uniform limit) color that marked in both figures corresponding to the profile of order parameter of cnoidal wave superflow in Fig.\ref{CW_bdry}. The black dashed line in Fig.\ref{jx_CW} lines out $\langle j^x \rangle=0$.}
\label{band}
\end {figure}

\textcolor{black}{The linear stability of the superflow solutions introduced in the previous section can be obtained by QNM analyses.} Just like the method proposed in \cite{{2016JHEP...01..016D,2020PhRvL.124c1601G,2020JHEP...02..104L,2021JHEP...11..190Y,2023arXiv230103203L}}, to calculate the QNM, we rewrite the equations of motions \eqref{A_mu} and \eqref{psi} in the infalling Eddington\textcolor{black}{-Finkelstein coordinates}:

\begin{equation}\label{Infalling Jx}
2 \partial_{t} \partial_{z}A_{x}-\partial_{z} \partial_{x}A_{t}-\partial_{z}\left(f \partial_{z}A_{x}\right)+i\left(\psi^{*} \partial_{x} \psi-\psi \partial_{x} \psi^{*}-2 i A_{x} \psi^{*} \psi\right)=0,
\end{equation}
\begin{equation}\label{Infalling Jz}
\begin{split}
\partial_{t} \partial_{z}A_{t}+\partial_{t} \partial_{x} A_{x}-\partial_{x}^{2} A_{t}-f \partial_{x} \partial_{z}A_{x}+i\left(\psi^{*} \partial_{t} \psi-\psi \partial_{t} \psi^{*}\right)\\
-if\left(\psi^{*} \partial_{z} \psi-\psi \partial_{z} \psi^{*}\right)+2 \psi^{*}A_{t} \psi=0,
\end{split}
\end{equation}
\begin{equation}\label{Infalling Jt}
\left(-\partial_{z}^{2}A_{t}+\partial_{x} \partial_{z} A_{x}\right)+i\left(\psi^{*} \partial_{z} \psi-\psi \partial_{z} \psi^{*}\right)=0,
\end{equation}
\begin{equation}\label{Infalling psi}
\begin{split}
2\partial_{t}\partial_{z}\psi-f \partial_{z}^{2}\psi-\left(f^{\prime}+2iA_{t}\right)\partial_{z}\psi-\partial_{x}^{2}\psi+2 i A_{x} \partial_{x}\psi\\
+\left(i\partial_{x}A_{x}-i\partial_{z}A_{t}+z+A_{x}^{2}\right) \psi=0,
\end{split}
\end{equation}
with the metric given by
$$ds^{2}=\frac{1}{z^{2}}\left(-f(z) d t^{2}-2dtdz+dx^{2}+dy^{2}\right).$$
Here and in the following sections we will not separate $\psi$ into real and imaginary parts. Similarly, there is one constraint equation, \textcolor{black}{which is chosen to be Eq.\eqref{Infalling Jz}}. Next, we give all the bulk field perturbations

\begin{equation}\label{eq:perturbation}
\begin{aligned}
A_t(t,z,x)=A^{BG}_t(t,z,x)+\delta A_{t}(t,z,x), \\
A_x(t,z,x)=A^{BG}_x(t,z,x)+\delta A_{x}(t,z,x), \\
\psi(t,z,x)=\psi^{BG}(t,z,x)+\delta \psi(t,z,x),
\end{aligned}
\end{equation}
where the ${A^{BG}_t, A^{BG}_x, \psi^{BG}}$ are the background fields and in our case they are the fields corresponding to superflow states. The continuous time translation symmetry and discrete space translation symmetry allow the perturbation fields with the following form
\begin{equation}\label{eq:perturbation}
\begin{aligned}
\delta A_{t}&=e^{-i \omega t+i q x} a(x)+e^{i \omega^{*} t-i q x} a^{*}(x), \\
\delta A_{x}&=e^{-i \omega t+i q x} b(x)+e^{i \omega^{*} t-i q x} b^{*}(x), \\
\delta \psi&=e^{-i \omega t+i q x} u(x)+e^{i \omega^{*} t-i q x}v^{*}(x).
\end{aligned}
\end{equation}
Substitute these perturbed fields into the equations of motion, we will get the linear perturbation equations
\begin{equation}\label{eq:perturbationEOMa}
\begin{array}{c}
\partial_{z}\left(\partial_{x} b+i q b\right)+\left(i v \partial_{z} \psi+i \psi^{*} \partial_{z} u\right)-\left(i u \partial_{z} \psi^{*}+i \psi \partial_{z} v\right)-\partial_{z}^{2} a=0,
\end{array}
\end{equation}
\begin{equation}\label{eq:perturbationEOMb}
\begin{array}{c}
\left(2 b \psi \psi^{*}+2\left(A_{x}-\frac{1}{2} q\right) \psi^{*}u+2\left(A_{x}+\frac{1}{2} q\right) \psi v\right)-2 i \omega \partial_{z} b+ \\
i\left(v \partial_{x} \psi+\psi^{*} \partial_{x} u\right)-i\left(u \partial_{x} \psi^{*}+\psi \partial_{x}v\right)-\partial_{z}\left(f \partial_{z} b\right)-\partial_{z}\left(\partial_{x}a+i q a\right)=0,
\end{array}
\end{equation}
\begin{equation}\label{eq:perturbationEOMu}
\begin{array}{c}
\left(z+\left(A_{x}-q\right)^{2}+i\left(\partial_{x} A_{x}-\partial_{z} A_{t}\right)\right) u+\left(2A_{x} b+i\left(\partial_{x} b+i q b-\partial_{z} a\right)\right) \psi- \\
\left(2 i \omega+f^{\prime}+2 i A_{t}\right) \partial_{z}u+2i\left(A_{x}-q\right) \partial_{x} u-f \partial_{z}^{2} u-\partial_{x}^{2} u-2 i a \partial_{z} \psi+2 ib\partial_{x} \psi=0,
\end{array}
\end{equation}
\begin{equation}\label{eq:perturbationEOMv}
\begin{array}{c}
\left(z+\left(A_{x}+q\right)^{2}-i\left(\partial_{x}A_{x}-\partial_{z} A_{t}\right)\right) v+\left(2 A_{x} b-i\left(\partial_{x}b+iqb-\partial_{z} a\right)\right) \psi^{*}- \\
\left(2 i \omega+f^{\prime}-2i A_{t}\right) \partial_{z} v-2 i\left(A_{x}+q\right) \partial_{x}v-f \partial_{z}^{2} v-\partial_{x}^{2} v+2 i a \partial_{z} \psi^{*}-2i b \partial_{x} \psi^{*}=0,
\end{array}
\end{equation}
where $\omega$ is the eigenvelue of the above linear equations and most of them have imaginary part $\omega_I$. The posetive value of $\omega_I$ means the latter time behavior of the superflow states under small perturbations will grow exponentially along with time, which is so-called the dynamical instability. In addition to that, if a unstable mode with negative value of $\omega_R$, which is real part of $\omega$, then there exsits the energetic (Landau) instability. In the follwing parts \ref{qnm}, we will see that the stability of cnoidal wave is different from the case in zero-temperature superfluid systems.

\begin{figure}
\centering
\begin{subfigure}[b]{0.3\textwidth}
\includegraphics[width = \textwidth]{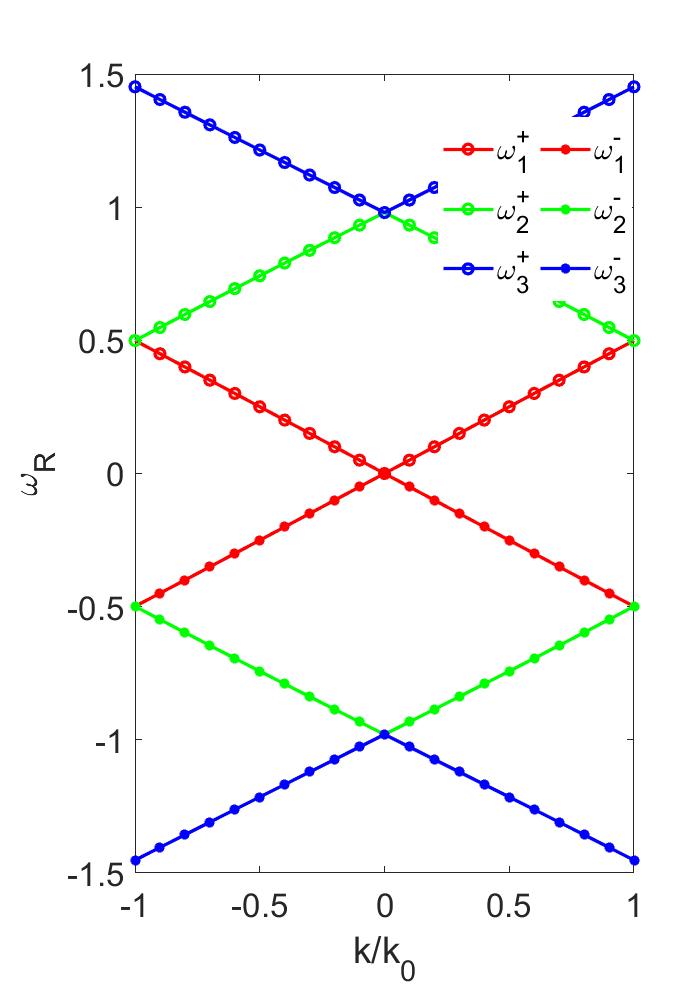}
\caption{}
\label{qnm_1}
\end{subfigure}
\begin{subfigure}[b]{0.3\textwidth}
\includegraphics[width = \textwidth]{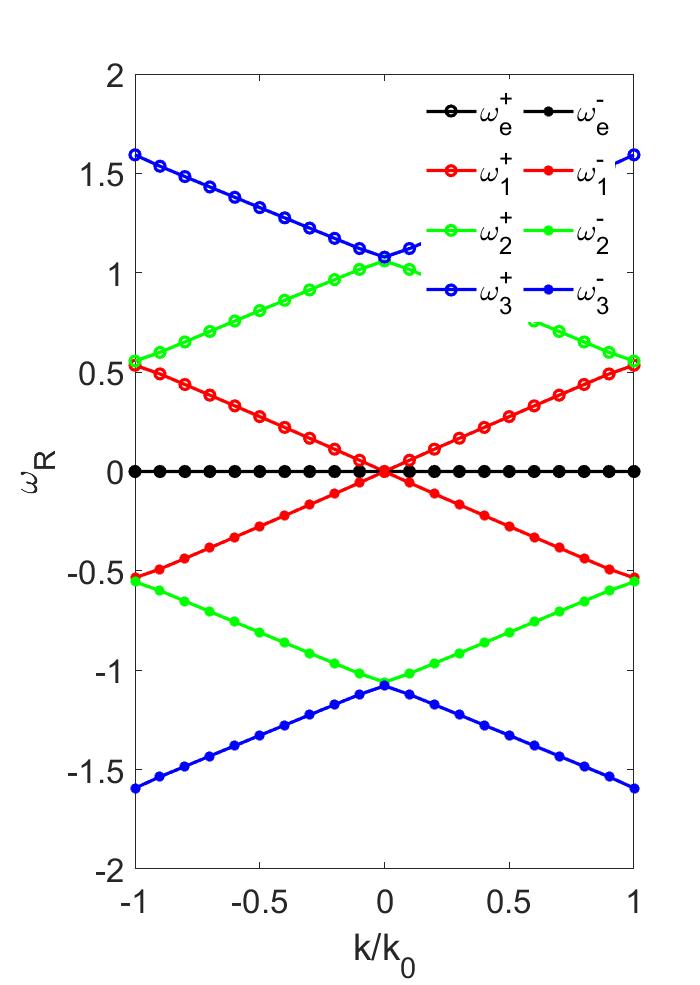}
\caption{}
\label{qnm_2}
\end{subfigure}
\begin{subfigure}[b]{0.3\textwidth}
\includegraphics[width = \textwidth]{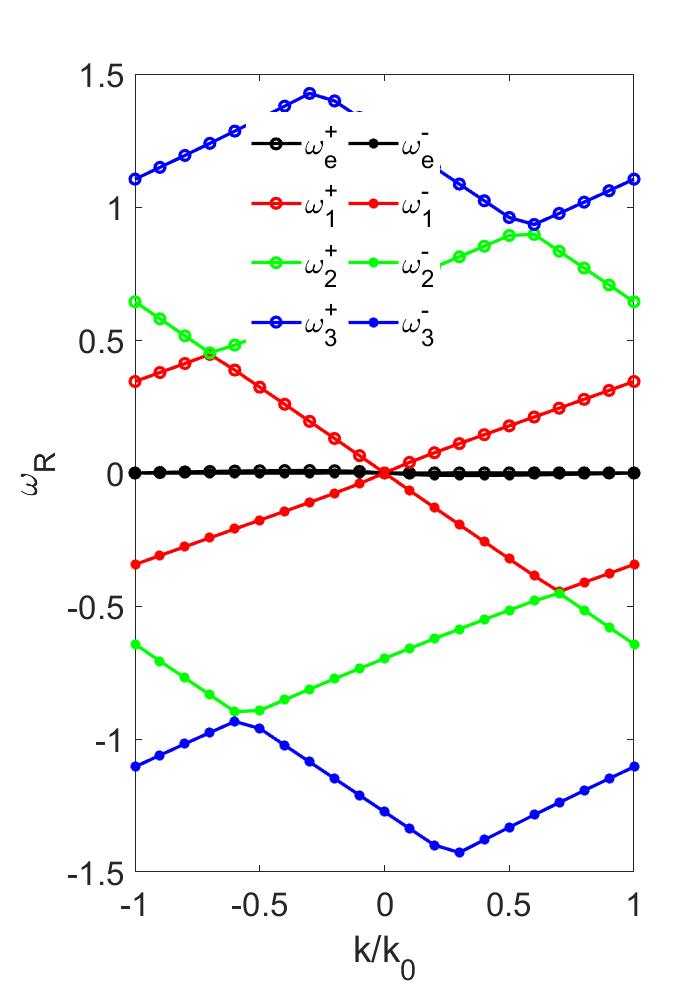}
\caption{}
\label{qnm_3}
\end{subfigure}
\begin{subfigure}[b]{0.45\textwidth}
\includegraphics[width = \textwidth]{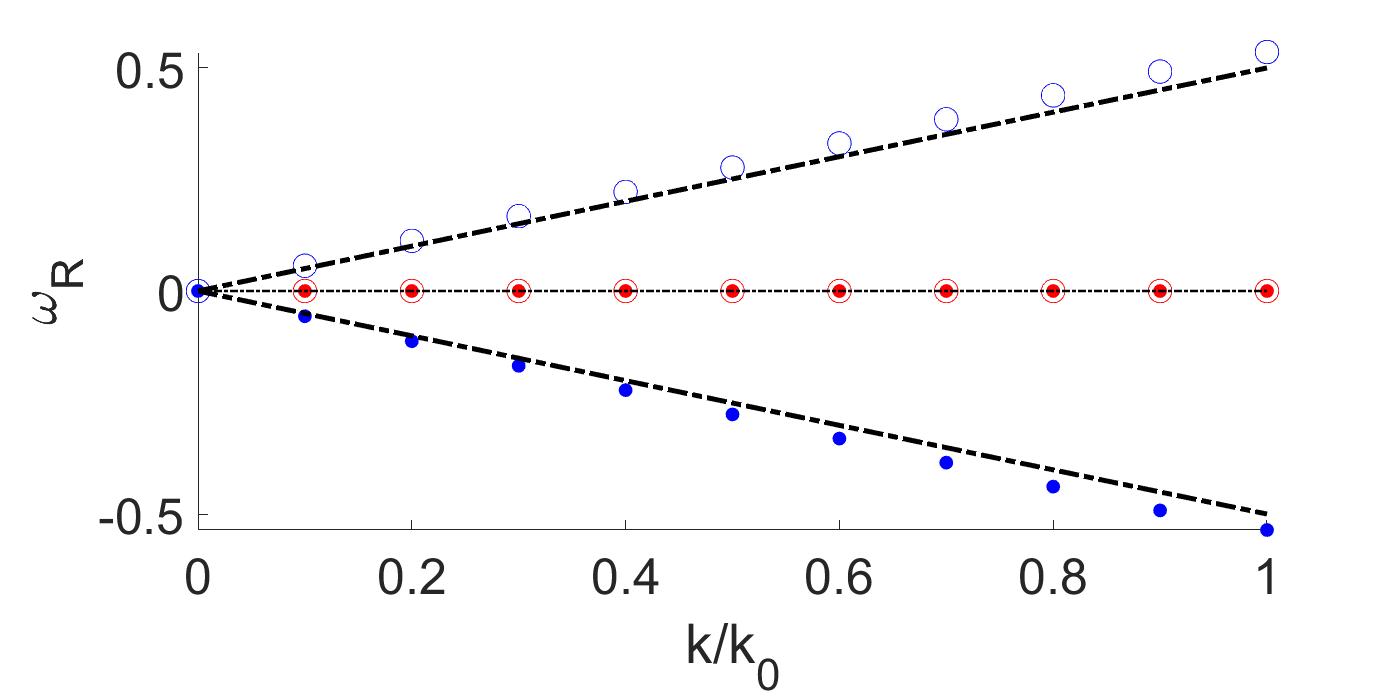}
\caption{}
\label{qnm_4}
\end{subfigure}
\begin{subfigure}[b]{0.45\textwidth}
\includegraphics[width = \textwidth]{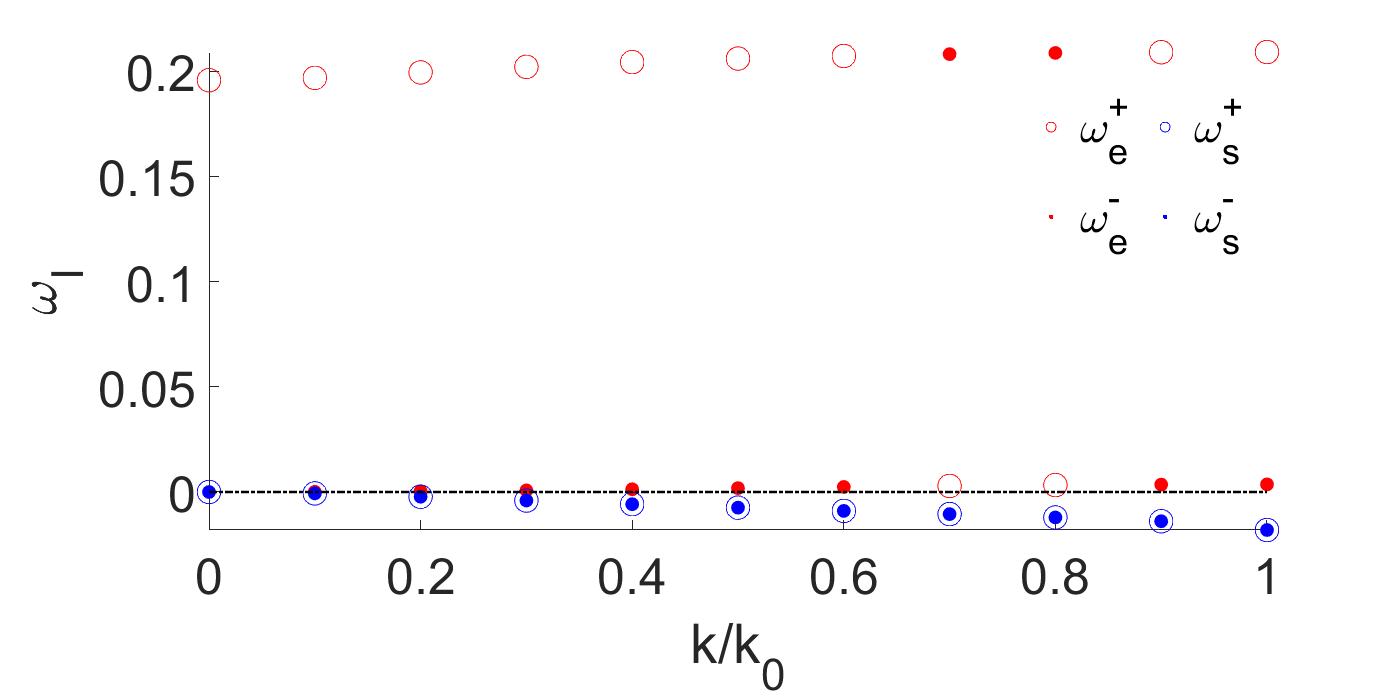}
\caption{}
\label{qnm_5}
\end{subfigure}
\begin{subfigure}[b]{0.45\textwidth}
\includegraphics[width = \textwidth]{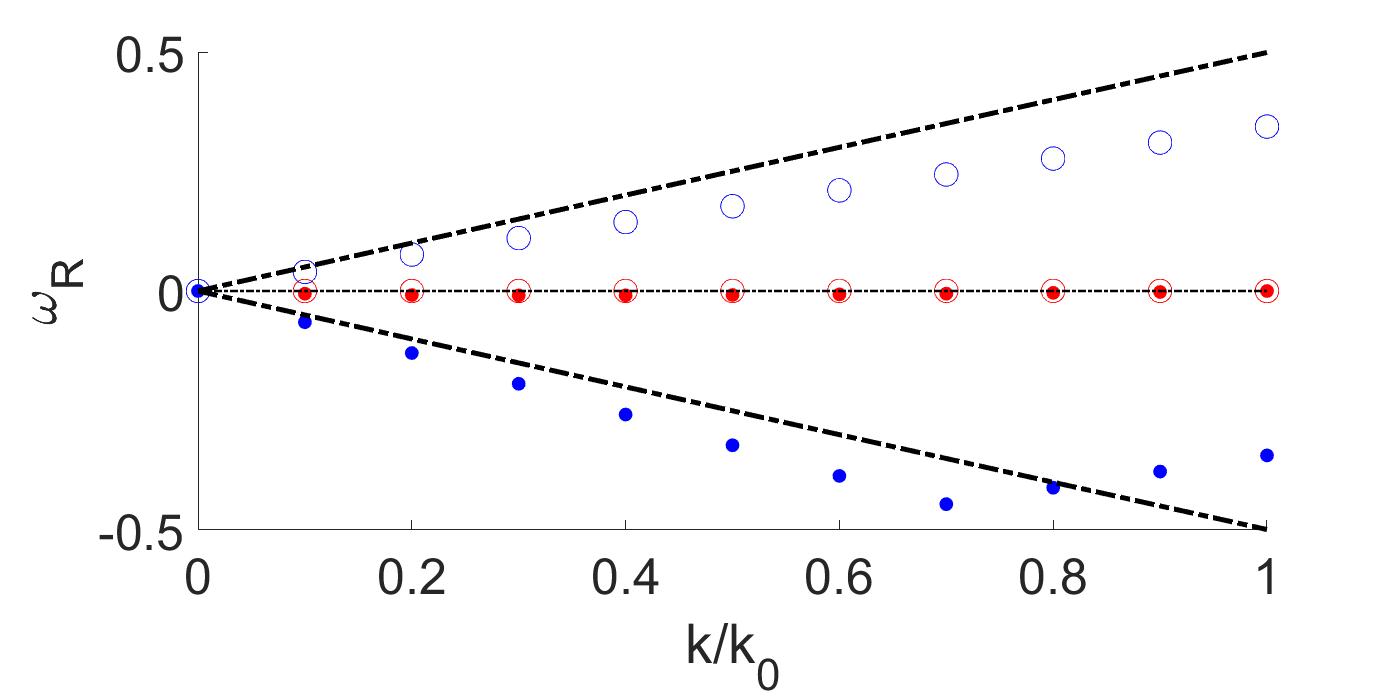}
\caption{}
\label{qnm_6}
\end{subfigure}
\begin{subfigure}[b]{0.45\textwidth}
\includegraphics[width = \textwidth]{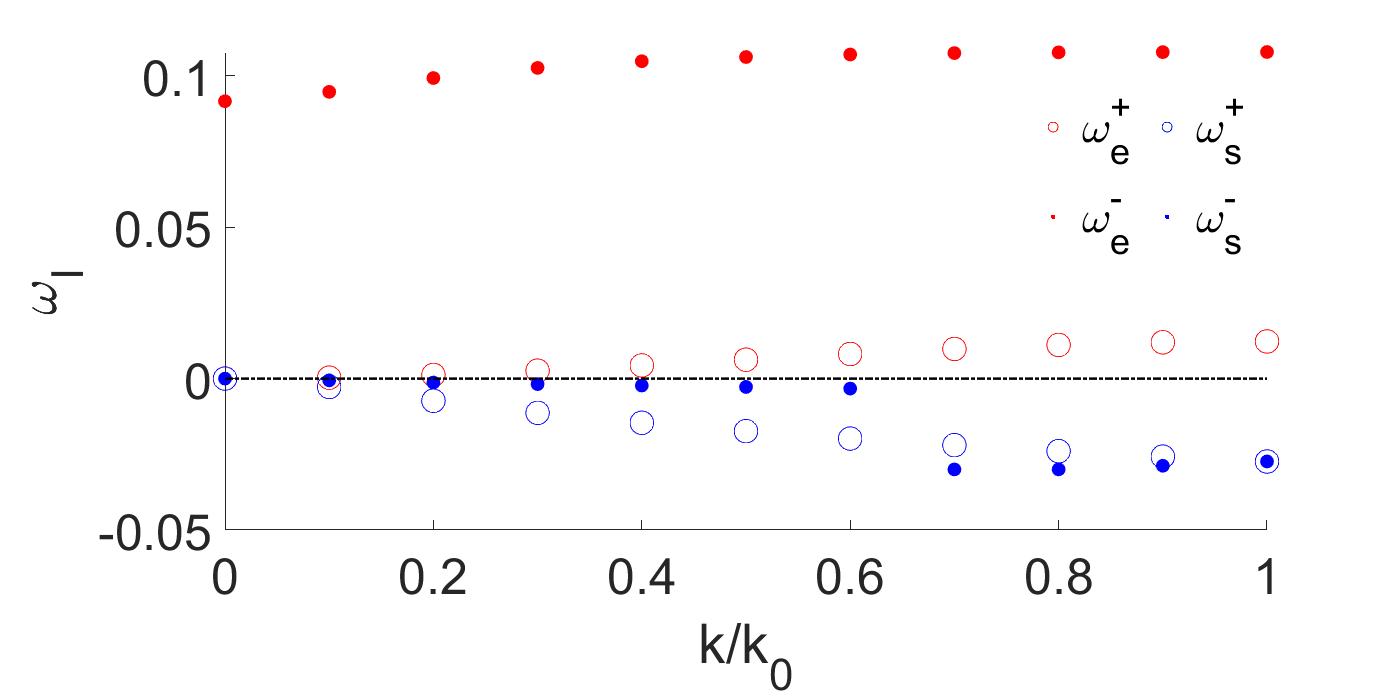}
\caption{}
\label{qnm_7}
\end{subfigure}
\caption{Fig.\ref{qnm_1} shows spectrum of $\omega_R$ for the homogeneous superflow ($a_x=0$). $\omega_i^{+}$ ($i=1,2,3$) are the modes with positive Re($\omega_i^{+}$) and $\omega_i^{-}$ are the modes with negative Re($\omega_i^{-}$). Fig.\ref{qnm_2} and Fig.\ref{qnm_3} show spectrum of $\omega_R$ for DSLCW and LWLCW, besides $\omega_i^{+}$ and $\omega_i^{-}$, there has additional elastic mode $\omega_e^{+}$ and $\omega_e^{-}$. Similarly, $\omega_e^{+}$ has positive Re($\omega_e^{+}$) and $\omega_e^{-}$ has negative Re($\omega_e^{-}$). One can see there have band gap in $\omega_i^{\pm}$ for both DSLCW and LWLCW, which can be confirm from the inexistence of overlapped points in neighbouring band. The details of real and imaginary parts for the two Goldstone modes ($\omega_e^{\pm}$ for elastic modes and $\omega_s^{\pm}$ for sound modes) of DSLCW and LWLCW are shown in Fig.\ref{qnm_4}-Fig.\ref{qnm_7}. The oblique thick black line in Fig.\ref{qnm_4} and Fig.\ref{qnm_6} are the real part of sound modes of the homogeneous superflow. With the help of these sound modes, we can distinguish $\omega_s^{\pm}$ and $\omega_e^{\pm}$ for DSLCW and LWLCW. The horizontal thin black line stands for the zero value, which are used to determine plus or minus characteristic of $\omega_e^{\pm}$ and $\omega_s^{\pm}$.
}
\end {figure}

\section{Numerical calculation and results}\label{result}
\subsection{Cnoidal waves}\label{CW}

To obtain the superflow states, besides the source of $A_{x}$ given by different value of $a_x$, we impose the source free boundary conditions for $\Psi_{0}$ at the conformal boundary, regular conditions for $\langle \mathcal O \rangle$ and $A_{t}=0$ at the horizon and periodic boundary conditions for $A_{t,x}$ and $\Psi$ in the $x$ direction. The situation we want to consider is particle number fixed superfluid system with different particle current density, so we will give a condition on the total particle number $N$ that defined as
\begin{equation}\label{N0}
N=\int^{\frac{L_x}{2}}_{-\frac{L_x}{2}} dx \rho(x),
\end{equation}
which is a kind of constrain on the chemical potential $\mu$ in $A_t$. We set spatial length $L_x=4$ and fix $N=N_0$ for all the calculation and the meaning of $N_0$ will be introduced latter. We solve static equations of motion numerically by the Newton-Raphson method and the solution we find are shown in Fig.\ref{states} with an example of bulk field $|\psi(z,x)|$ in Fig.\ref{CW_bulk} and the profile of $|\langle \mathcal O(x) \rangle|$ of three different limiting in Fig.\ref{CW_bdry}.

Since there will be uniform limiting case for cniodal wave, we also solve homogeneous superflow with the same source $a_x$. The results are shown in Fig. \ref{band}. To solve homogeneous superflow with $a_x=0$, instead of fixing total particle number, we fix the chemical potential $\mu=6$. Finally, we obtain total particle number in this state via Eq.\eqref{N0}, which is $N_0$ we mentioned before. From Fig. \ref{band}, one can see that as the source $a_x$ increasing to $\pi$, both the chemical potential $\mu$ and particle number density $\langle j^x \rangle$ become the same value. But before that, the cnoidal wave and homogeneous superflow are two different solutions. From Fig.\ref{Band_CW}, there is a minimum in $\mu(a_x)$ at $a_x=\frac{pi}{4}$ (marked as red filled triangle), which corresponding to $\langle j^x \rangle=0$ in Fig.\ref{jx_CW}.  When $a_x$ closed to $\frac{pi}{4}$, the cnoidal wave are soliton train with constant particle number current. As $a_x$ becomes larger but not reaches $\pi$, it corresponding to linear-wave limit case and the profile of $|\langle \mathcal O(x) \rangle|$ is shown in Fig.\ref{CW_bdry}. The larger chemical potential $\mu(a_x)$ of cnoidal wave gives us a clue that the cnoidal wave is unstable, but in the next section we will talk about the dynamical stability and energetic stability  of cnoidal wave more clearly.


\subsection{Spectrum of QNM and instability of cnoidal waves}\label{qnm}
Without loss of generality, we consider two special cases, the dark-soliton limiting cnoidal wave ($a_x=\frac{pi}{4}$ and we donate as DSLCW) with zero particle current density and linear-wave limiting cnoidal wave ($a_x=\frac{7 pi}{8}$ and we donate as LWLCW) non-zero particle current density. Due to the spatial crystalline order, these spectrum of $\omega_R$ should have band structure. In Fig.\ref{qnm_2} and Fig.\ref{qnm_3}, we respectively show the spectrum of $\omega_R$ for DSLCW and LWLCW and the spectrum of $\omega_R$  for the homogeneous superflow ($a_x=0$) is shown in Fig.\ref{qnm_1} as a comparison. There are two mainly differences from Fig.\ref{qnm_1} to Fig.\ref{qnm_2} and Fig.\ref{qnm_3}, the first is there has only one Goldstone mode for homogeneous superflow and two Goldstone modes for the latter, which is the fundamental property of supersolids; the second is for the DSLCW and LWLCW the band structure of spectrum of $\omega_R$ are shown clearly, which is also an affirmation for the supersolidity.

The stability of cnoidal waves are embedded in the Goldstone modes, so in Fig.\ref{qnm_4}-Fig.\ref{qnm_7}, we show the details of the sound mode and the elastic mode for DSLCW (\ref{qnm_4} and Fig.\ref{qnm_5}) and LWLCW (\ref{qnm_6} and Fig.\ref{qnm_7}). From $\omega_I$ we can see, both of the DSLCW and LWLCW have dynamical instability due to the elastic mode $\omega_e^{-}$. For the DSLCW, the real part of $\omega_e^{+}$ and $\omega_e^{-}$ are degenerate, i.e., $Re(\omega_e^{+}(k))=Re(\omega_e^{-}(k))=0$ for all the $k$, which means the DSLCW is energetic stable. However, for the LWLCW, the non-degenerate of Re$(\omega_e^{+})$ and Re$(\omega_e^{-})$ will lead to the unstable $\omega_e^{-}$ mode with Re($\omega_e^{-})<0$ and Im($\omega_e^{-})>0$, which stands for the LWLCW has both the dynamical instability and energetic instability at the same time.

\section{Summary}\label{sum}
The cnoidal waves solutions are obtained in finite temperature holographic superfluid system numerically and three different limiting cases for conidal waves (dark-soliton limit, linear-wave limit and uniform limit) can be easily found by introducing the source $a_x$ for particle current density when solving the equation of motions. When we increase $a_x$, the cnoidal waves will from dark-soliton limit passing linear-wave limit to uniform limit eventually. And both of dark-soliton limit cnoidal wave and linear-wave limit cnoidal wave have spatial crystalline order spontaneously, which lead to the supersolidity-related phenomena of these cnoidal waves. The value of chemical potential $\mu$ and particle current density $\langle j^x \rangle$ will have the same value with homogeneous superflow after cnoidal waves enter into uniform limit. 

To investigate the fundamental supersolidity of cnoidal waves, we take some quasi-normal mode analysis for the dark-soliton limiting cnoidal wave and linear-wave limiting cnoidal wave. The appearance of two Goldstone modes, sound mode as well as elastic mode, and the band structure of spectrum of quasi-normal modes confirm that the $U(1)$ symmetry and continuous space translation symmetry are broken spontaneously. Different from the zero temperature cnoidal waves, the finite temperature cnoidal waves will always have dynamical instability. Besides, when $\langle j^x \rangle \neq 0$, the energetic instability will also company with dynamical instability. And these instability are both from the elastic modes $\omega_e^{-}$.

\section*{Acknowledgement}
This work is partially supported by the Natural Science Foundation of China under Grants No. 11975235 and No. 12035016.







\end{document}